\documentclass[prd,twocolumn,showpacs,preprintnumbers]{revtex4}
\usepackage{amssymb}
\usepackage{graphicx}
\usepackage{dcolumn}
\usepackage{bm}
\usepackage{amsfonts}
\usepackage[russian]{babel}

\newcommand{\la}{\lambda}
\newcommand{\lb}{\label}
\newcommand{\var}{\varepsilon}
\newcommand{\vE}{{\bf{E}}}
\newcommand{\vl}{v_{||}}
\newcommand{\vpp}{v_{\perp}}
\newcommand{\om}{\omega}
\newcommand{\al}{\alpha}

\newcommand{\pa}{\partial}

\newcommand{\fr}{\frac}
\newcommand{\de}{\delta}
\newcommand{\bw}{\begin{widetext}}
\newcommand{\ew}{\end{widetext}}
\newcommand{\be}{\begin{equation}}
\newcommand{\ee}{\end{equation}}
\newcommand{\bee}{\begin{equation*}}
\newcommand{\eee}{\end{equation*}}
\newcommand{\ba}{\begin{eqnarray}}
\newcommand{\ea}{\end{eqnarray}}
\newcommand{\bal}{\begin{align}}
\newcommand{\eal}{\end{align}}
\newcommand{\bml}{\begin{multline}}
\newcommand{\eml}{\end{multline}}
\newcommand{\non}{\nonumber}

\newcommand{\vn}{{\bf n}}

\newcommand{\vk}{{\bf k}}

\newcommand{\vp}{{\bf{p}}}
\newcommand{\vB}{{\bf{B}}}

\newcommand{\vev}{\bf{v}}

\newcommand{\Nv}{{\mathbf{N}}}

\def\cF{{\cal F}}
\def\il{{\it{ l}}}

\def\vk{{\bf k}}
\def\vr{{\bf r}}
\def\cF{{\cal F}}

\def\e{{\rm e}}

\begin{document}
\title{Plasmon-graviton conversion in a magnetic field in TeV-scale
gravity.}
\author{ E.Yu. Melkumova\footnote{email: elenamelk@srd.sinp.msu.ru}}
\affiliation{Moscow State University, 119899, Moscow Russia}
\pacs{11.25Wx, 11.27+d}
\date{\today}
\begin{abstract}
Kaluza-Klein (KK) gravitons emission rates  due to plasmon-graviton
conversion in magnetic field are computed within  the ADD model of
TeV-scale gravity. Plasma is described in the kinetic approach as
the system of charged particles and Maxwell field both confined on
the brane. Interaction with multidimensional gravity living in the
bulk with $n$ compact extra dimensions is introduced within the
linearized theory. Plasma collective effects enter through  the
two-point correlation function of the fluctuations of the
energy-momentum tensor. The estimate for magnetic stars is presented
leading to the lower limit of the D-dimensional Plank mass.
\end{abstract}
\maketitle
\section{Introduction}
Photon-graviton conversion  in magnetic  field in four-dimensional
space-time \cite{G62} was discussed earlier in a number of papers
including the monograph \cite{ZN83}. It was argued that plasma
effects will make the photon (plasmon) -- graviton conversion
process astrophysically negligible \cite{CH96}, because the
characteristic length of photon-graviton oscillations
 $${\it l}_{\rm osc}=\fr{4\pi\om}{\om_L^2}\ll L,\quad {\rm
 where}\quad
\om_L=\sqrt{\fr{4\pi^2 e^2n}{m}} $$ is the plasma electron Langmuir
frequency will be
 much smaller then typical homogeneity distance  of
 magnetic field ${B}$. In fact, the resonance is destroyed since the dispersion
 relation  for the graviton is $\om=k$, while for the plasmon
 $ \om^2 = k^2+\om_L^2.$
 \par
 Recently, a lot of interest has been attracted to models with large
    extra-dimensions
  \cite{ADD99,ADD98A}. According to one such scenario, called ADD,
the standard model particles live in the four-dimensional subspace
(the brane) of the D-dimensional bulk with $ n = D - 4 $ extra
dimensions compactified on a torus which are inhabited only by
gravity. The D-dimensional Planck mass $M_D$ is supposed to be
TeV-scale, and the large extra dimensions (LED) to have
sub-millimeter size. In this scenario there are KK gravitons which
are seen in four dimensions as a tower of massive particles  (the
mass $M$) with the dispersion relation $ \om^2=k^2+M^2 $. This can
restore the resonance condition for plasmon-graviton conversion in
magnetic field if $\om_L=M$.

\section{Self-consistent kinetic theory for magnetized plasma}
Consider collisionless plasma consisting of charged particles of
types $\al$ with charges and masses $e_\al,\,m_\al$ described by the
microscopic distribution function \cite{Sit67,GGM83}
 \be
\cF_\al(x,\vp)=\sum_{i=1}^{N_\al}\de(\vr-\vr_i(t))\de(\vp - \vp_i
(t)),\label{mdf}
  \ee
 normalized as
 $  \int\cF_\al(x,\vp)d{\bf p} \,\,d^3 x=N_\al, $ and satisfying the
 kinetic equation
 \be
 \fr{\pa \cF_\al}{\pa t}+\vev
\fr{\pa \cF_\al}{\pa \vr} + e_\al
(\bf{E}+[\vev\bf{B}]+[\vev\bf{B}_0])\fr{\pa \cF_\al}{\pa \vp}=0.
\lb{cf}
 \ee
They move in the magnetostatic field $\vB_0$ and interact via
electromagnetic fluctuating field $\bf{E},\,\bf{B}$ satisfying the
Maxwell equations
    \ba && {\rm div} \vE=4\pi\rho, \quad
 {\rm rot}\vE=-\fr{\pa \vB}{\pa t},\non\\&&
{\rm div}\vB=0,\quad {\rm rot}\vB =4\pi\bf{j} +\fr{\pa \vE}{\pa
t}, \lb{max}\ea with the source terms
 \ba \label{sorce}&&\rho(x)=\sum_\al
e_\al\int\cF_\al(x,{\vp})d{\bf p},\non\\&&
 {\bf{j}}(x)=\sum_\al
e_\al\int\fr{{\bf p}}{p_\al^0}\cF_\al(x,{\vp})d{\vp}.
 \ea
To solve the system of equations (\ref{cf}-\ref{sorce})  we use
perturbation theory in terms of the electric charges. First we
separate fluctuations from the average distribution using the
approach of
 \cite{Sit67}:
\be \cF(t,{\vr},{\vp})=f^0_\al+\de f_{\al}^0+\de f_\al,\lb{f}\ee
 where
$f^0\equiv <\cF_\al(x,{\vp})>$ - the  equilibrium Maxwell
distribution function: \ba && f_\al^0=N_{0\,\al}\left(\fr{m}{2\pi
 T_{\al\,\perp}}\right)^{\fr32}
 \sqrt{\fr{T_{\al\,||}}{T_{\al\,\perp}}}\e^{-\fr{ m_\al\vpp^2}{2T_{\al\,\perp}}} \e^{-\fr{ m\vl^2}{2T_{\al\,||}}} ,
 \non\\&&
 v_{T_{\al\,\perp}}=\sqrt{\fr{T_{\al\,\perp}}{m_\al}},
 v_{T_{\al\,||}}=\sqrt{\fr{T_{\al\,||}}{m_\al}}<<1, \lb{max}\ea
where the symbols $\perp$ and $||$ are introduced for transversal
and longitudinal thermal velocities with respect to the
magnetostatic field.
 Here $\de f^0_\al$- represents fluctuations due to chaotic particle motion
("zero" fluctuations), while $\de f_\al(t,{\vr},{\vp})$ - stands for
fluctuations arising due to  their electromagnetic interaction.
Using the fact that $f_\al^0$ satisfies the free kinetic equation,
we obtain: \ba \!\!\fr{\pa \de f_\al}{\pa t}+{\vev }\fr{\pa \de
f_\al}{\pa \vr}\, +\, e_\al {{(\bf{E}\!+\![\vev\bf{B}])}}\fr{\pa (
f^0_\al+\de f_{\al}^0+\de f_\al )}{\pa \vp}\!=\!0. \lb{cfg}\ea
Using the Fourier-transformation
$$ f(x)=\fr{1}{(2\pi)^4}\int f(k)\e^{-ik_\mu x^\mu}d^4k, $$ the solution of the self-consistent system  equations
(\ref{cf}-\ref{sorce},\ref{cfg}) in linear approximation is well
known \cite{Sit67}-\cite{GR72}.
\par
Maxwell equation may be written as \ba\lb{lmax}&&
\Lambda_{ij}E_j=-\fr{4\pi i}{\om} j^0_{i}, \non\\&&
{\bf{j}}^0(x)=\sum_\al e_\al\int{\vev}\de f^0_\al(x,\vp)d^3p .\ea
 Here \be
 \lb{Lam}\Lambda_{ij}=\var_{ij}-\fr{k^2}{\om^2}(\de_{ij}-\hat{k}_i\hat{k}_j),
\ee where
 $\var_{ij}$ is the permittivity tensor, $ \hat{\vk}=\fr{\vk}{|\vk|} $, $ \bf{j}^0 $ - "zero" fluctuation current density.\\

 For cold plasma in magnetic field satisfying the conditions
  \ba\lb{condk}  \fr{k_\perp v_{T_\al}}{\om_{B_{\al}}}\ll 1,\quad \fr{k_z v_{T_\al}}{\om}\ll
  1,\quad
  \fr{\om\pm \om_{B_{\al}}}{k_z v_{T_\al}}\gg 1. \ea
 the permittivity tensor may be presented as:
  \[\lb{cpl} \var_{ij}=\left(\begin{array}{lcr}
\var_\perp & i g & 0 \\
-i g  &  \var_\perp & 0
 \\
0 & 0 & \var_{||}
\end{array} \right),\]
where \ba\lb{ecp}&& \var_\perp =
\var_{xx}=\var_{yy}=1-\sum_\al\fr{\om^2_{L_{\al}}}{\om^2
-\om^2_{B_{\al}}},\non\\&&
 \var_{||}=\var_{zz}=1-
\sum_\al\fr{\om^2_{L_{\al}}}{\om^2},\non\\&&
\var_{xy}=-\var_{yx}=ig=-i\sum_\al\fr{\om^2_{L_{\al}}\om_{B_{\al}}}{\om(\om^2
-\om^2_{B_{\al}})},\non\\&&
\var_{xz}=\var_{zx}=\var_{yz}=\var_{zy}=0.\ea The average  of
 fluctuational field will be
zero, while the quadratic correlation functions are non-zero
\cite{Sit73}.
 Fluctuations can be characterized by correlation functions whose Fourier-transforms
 exhibit
 homogeneity, e.g.
 \ba && <j^0_i(\vr_1,t_1)j^0_j(\vr_2,t_2)>\equiv
 <j_ij_j>^0_{\vr,t},\non\\&&
 \vr=\vr_1-\vr_2,t=t_1-t_2.\ea
 The space-time Fourier transformation will give the spectra of current fluctuations:
  \ba &&
  <j_ij_j>^0_{\vk\om}=\int d\vr\int d t
  \e^{-i\vk\vr+i\om\vr}<j_ij_j>^0_{\vr,t},\non\\&&
 <j^0_i(\vk,\om)j^0_j(\vk',\om')>
 =(2\pi)^4<j_ij_j>^0_{\vk\om}\cdot\non\\&&
 \de(\om - \om')\de(\vk -
 \vk').\ea
Correlation function for "zero" current  $<j_ij_j>^0_{\vk\om }$
plays the most important role in the magnetized plasma.  For the
Maxwell distribution function (\ref{max}) it reads:
 \ba\lb{c8} {<j_ij_j>^0_{\vk\om }}={2\pi e_\al^2\int d^3v \sum_n
  \Pi_{ij}\de(\om -k_{||}v_{||}-\il \om_B )}f_0(\vev) \non\ea
\[ \Pi_{ij}=\left[\begin{array}{lcr}
v_\perp^2\fr{\il^2}{b_\al^2}J_\il^2(b_\al) & i v_\perp^2
\il\fr{J_\il J_\il'}{b_\al} & v_\perp v_z \il \fr{J_\il^2}{b_\al} \\
-i v_\perp^2 \il\fr{J_\il J_\il'}{b_\al} & v_\perp^2J_\il'^2  &
-i v_\perp
v_z J_\il J_\il' \\
v_\perp v_z \il \fr{J_\il^2}{b_\al} &  i v_\perp v_z J_\il J_\il'
& v_z^2 J_\il^2
\end{array} \right] ,  \]
 where $b_\al=\fr{k_\perp v_\perp}{\om_{B_{0\al}}}.$
Correlation functions for fluctuational fields may be expressed in
terms of zero current correlation function. Due to the equations
(\ref{lmax}), the following relation holds
  between the spectral functions $<j_ij_j>^0_{\vk\om}$  and
  $<E_iE_j>_{\vk\om}$:
\ba\lb{eiej}&&
<E_iE_{j}>=\fr{16\pi^2}{\om^2}\fr{\la_{ip}\la^*_{jp'}}{\Lambda^2}<j_pj_{p'}>^0,
\non\\&& \fr{\la_{ip}}{\Lambda}=\Lambda_{ip}^{-1} ,\quad
\Lambda^{-1}_{ip}\Lambda_{is}=\de_{ps}, \ea where $ \la_{ij} $
-algebraic adjunct of matrix $\Lambda_{ij}$, $
\Lambda=\rm{det}|\Lambda_{ij}|. $
\section{Plasmon-graviton transformation}
Formula for the total (integrated over space and time) energy loss
on gravitational radiation in ADD model for linearized $D=4+d$
dimensional Einstein equations was recently derived in
\cite{GKST10}. The  corresponding expression for the gravitational
emission rate to all graviton KK modes per unit plasma volume per
unit time in the stochastic plasma system  was given in
\cite{M11}: \ba && P=\frac{\varkappa_D^2}{16 \pi^{3} V_d}\sum_{\Nv
\in \mathbb{Z}^d} \int d{\vk} <\de T_{ik} \de
T^*_{i'k'}>_k\tilde{\Lambda}^{iki'k'} \left. \vphantom{\sqrt{d}}
    \right|_{\om=k^0},\non\\&&
    k^0=\sqrt{|{\vk} |^2+M^2},\quad V_d=(2\pi L)^{D-4},
\lb{lastADD2}\\&&
\tilde{\Lambda}^{iki'k'}=\frac{1}{2}\left[\Delta^{ii'} \Delta^{
kk'}+\Delta^{ik'} \Delta^{ i'k}\right]-\frac{1}{D-2}\Delta^{ik}
\Delta^{ i'k'} , \non\\&&
 \Delta_{ik}=\de_{ik}-n_in_k, \quad
{\bf{n}}=\fr{{\vk}}{{k^0}},\quad M^2=(2 \pi \Nv /L)^2,\lb{LA}\ea
where $L$ is the size of the extra dimensions, $\varkappa_D^2=16\pi
G_D,\quad G_D=M_D^{-(d+2)}$. In our approach we did not use explicit
decomposition of the full multidimensional metric perturbation
$h_{MN}$ in massless and massive modes, these correspond to $\Nv=0$
and $\Nv\neq 0$ terms in the sum respectively. For massless modes
the three-dimensional vector ${\bf{n}}$ is the unit vector, but not
for massive modes. Note that $h_{MN}$ has bulk components due to the
trace term in the wave equation.
\par
  The gravitational radiation due to
 plasmon-graviton transformation possess  will be generated by the field part
of the energy-momentum tensor: \ba\lb{ft}&& {}_F
T_{ij}=-\fr{1}{4\pi} \left( B_iB_{0\,j}+B_{0\,i}B_j-\de_{ij}{(\vB
\vB_0)}\right),\non\\&&
  B_i
=\fr{1}{\om}\epsilon_{ijk}k_jE_k, \ea where $\vB,\vE$ are the first
order fluctuation fields, $\epsilon_{ijk}$ is unit antisymmetric
tensor.

 After substitution of (\ref{ft})  into the expression for the radiation
power and contraction with the projector $\Lambda_{iki'k'}$ the
radiation per unit time per unit volume may be presented as
\ba\lb{gP1}&& P =\frac{\varkappa_D^2}{8 \pi^{3} V_d}\sum_{\Nv \in
\mathbb{Z}^d} \int  d\vk n^2\Big[(B_0^2 -(\vn
B_0)^2)(\de_{ss'}-\hat{k}_s\hat{k}_{s'})+\non\\&&
\fr12\fr{(D-4)[(n^2-1)^2+8]+[(n^2+3)^2-16]}{D-2}\cdot\non\\&&
[\vB_0\hat{k}]_s[\vB_0\hat{k}]_{s'}\Big] <E_sE_{s'}>, \quad
\hat{\vk}=\fr{\vk}{k}.\ea
 \subsubsection{Transversal plasmon $\om^2=k^2+\om_L^2$.}
 The detailed dispersion relation in a strongly magnetized plasma
is more complicated \cite{GR72} than   the simple relation
$\om^2=k^2+\om_L^2$, that was used in the early papers on this
subject \cite{RS88},\cite{DU00}.
 The condition of existence of plasma waves (coherent
fluctuations) is  $\quad Re \Lambda \, =0\,\quad .$
  In   magnetized "cold" plasma
 the
dispersion relation for the coherent fluctuations are resolved
analytically only for certain sites of a spectrum, at certain
direction   with respect to the magnetic  field and in general
there
 are not strictly longitudinal or transversal waves.
 The ordinary
 transverse  high frequency electron wave  $ \om^2=k^2+\om_L^2 $
 propagates
 {\em{strongly
 across}} the magnetic field: {\ba\lb{cond} &&\om^2=k^2+\om_L^2,
 \quad k_z=0,\quad k_\perp=k  ,
    \non\\&&
      k_z^2\ll \fr{\om^6 (\om^2 -\om_B^2)}{\om_L^4 \om_B^2}.\ea }
The field correlation function for coherent fluctuation may be
rewritten as \cite{Sit67}
 \ba\lb{eiej1}
<E_iE_{j}>=\fr{16\pi^3}{\om^2}\fr{\la_{ip}\la_{jp'}\de
(\rm{Re}\Lambda)}{| \rm{Im}\Lambda |}<j_pj_p'>^0 ,\ea where $ |
\rm{Im}\Lambda | $ is determined by the small thermal terms of the
anti-hermitean part of the permittivity tensor
   \ba\lb{ecp1}&&
   \de \var^a_{xx}\equiv\de
\var^a_\perp,\non\\&&
\de
\var^a_\perp=i\sqrt{\fr{\pi}{8}}\sum_\al\fr{\om_{L_\al}^2}{\om
|k_z |v_{T_\al}} \left[ \e^{-\fr{(\om-\om_{B_\al})^2}{2k_z^2
v_{T_\al}}}+ \e^{-\fr{(\om+\om_{B_\al})^2}{2k_z^2
v_{T_\al}}}\right],\non \\&&
 \de \var^a_{yy}=\de \var^a_\perp
+i\sqrt{2\pi}\sum_\al\fr{\om_{L_\al}^2k^2_\perp
v^2_{T_\al}\om}{\om |k_z |v_{T_\al}\om_{B_\al}^2}
\e^{-\fr{\om^2}{2k_z^2 v_{T_\al}}},\non\\&&
 \de
g^a=i\sqrt{\fr{\pi}{8}}\sum_\al\fr{\om_{L_\al}^2}{\om |k_z
|v_{T_\al}} \left[ \e^{-\fr{(\om-\om_{B_\al})^2}{2k_z^2
v_{T_\al}}}+ \e^{-\fr{(\om+\om_{B_\al})^2}{2k_z^2
v_{T_\al}}}\right],\non \\&&
 \de\var_{||}=
i\sqrt{\fr{\pi}{8}}\sum_\al\fr{\om\om_{L_\al}^2}{ k_z^3
v_{T_\al}^3}\e^{-\fr{\om^2}{2k_z^2 v_{T_\al}}}. \ea According to
the condition (\ref{cond}), the imaginary part of the determinant
of the Maxwell tensor is equal to \ba\lb{ip}  \rm{Im}\Lambda
=\left( \var_\perp^2 -g^2-\fr{k^2}{\om^2}\var_\perp
\right)\de\var_{||}, \ea and the only   nonzero term of the
correlation function is \ba\lb{st} <j_zj_z>^0= \fr{4 e^2 N_0}{\pi
m}
 \sqrt{T_{||}T_\perp}\fr{\om}{\om_{L_e}^2}\de\var_{||},\ea
 and single term for adjunct tensor $\la_{sp}$ is \ba\lb{lzz} \la_{zz}=\var_\perp^2 -g^2-\fr{k^2}{\om^2}\var_\perp
 =\fr{\om_L^2\om_B^2}{\om^2(\om^2-\om_B^2)} .\ea
The delta-function entering the expression (\ref{eiej1}) for the
 transverse plasmon with $\om^2=k^2+\om_L^2$ may be presented as
 \ba\lb{rl} \de(\rm{Re}\Lambda) =\fr{\om^4(\om^2-\om_B^2)}{\om_B^2\om_L^2}|_{\om^2=k^2+\om_L^2}\de(\om^2-k^2-\om^2_L).\ea
 Radiation rate(\ref{gP1})  for the transformation possess may be
 then rewritten as
  \ba\lb{gP2} P \!=\!\frac{2\varkappa_D^2B_0^2{\sqrt{T_\perp
T_{||}}}}{\pi^{2} V_d}\!\!\!\sum_{\Nv \in \mathbb{Z}^d} \!\!\int
\!\!\!d\vk n^2\!\!\sqrt{k^2+\om_L^2}\de(M^2-\om_L^2). \ea The next
step is to sum over the   KK modes. Assuming that a large number
of modes is excited, one can replace the summation over $N$ by
integration and integrating over $k_z$  taking into account the
condition (\ref{cond}) we obtain: \ba\lb{gP3}&& P
=\frac{2\varkappa_D^2\pi^{\left(\fr{d}{2}-1\right)}}{\Gamma\left(\fr{d}{2}\right)
} \fr{B_0^2\sqrt{T_\perp T_{||}}\om_L^{(d-4)}}{\om_B}\cdot\non\\&&
\int\!\!dk {k^3(k^2+\om_L^2)\ }{\sqrt{k^2+\om_L^2-\om_B^2}}\,. \ea
Here according to the conditions $ \fr{e^2}{r_{min}} < T $ we
introduced the cutoff parameter $k_{max}=\fr{1}{r_{min}}$.
\subsubsection{Astrophysical estimates.} After integration over
$k$ let us rewrite expression (\ref{gP3}) in the CGSE system of
units and for an estimate assume that $T_\perp\simeq T_{||}\equiv T$
:
 \ba\lb{gPlast}
P\simeq\fr67\frac{20^{13}\pi^{\left(\fr{d}{2}+7\right)}}{\Gamma\left(\fr{d}{2}\right)
} \fr{B_0^2\om_L^{(d-4)}}{M^{d+2}\om_B}. \fr{T^8}{\al^7} \quad
\fr{\rm{erg}}{\rm{cm}^3 \rm{s}}.\ea Here $\al=\fr{1}{137}$, and the
Langmuir electron frequency $\om_L$, the cyclotron frequency
$\om_B$, the electron temperature $T$  and the D-dimensional Plank
mass $M_D$ have to be expressed in eV.
\par
In the ADD scenario the  massive Kaluza- Klein (KK) gravitons
contribute to energy losses of astrophysical objects
\cite{BHKZ99}-\cite{HR03} leading to bounds on $M_D$. Let us
estimate contribution of plasmon-graviton transformation in neutron
stars using data  \cite{ZPS96} for millisecond pulsar J0437-4715: $
B=10^9\,\rm{gauss},\quad N\sim10^{24}\rm{cm}^{-3},\quad T=10^{5.6}
K. $ The conservative upper limit of the energy-loss rate of
magnetic neutron stars $$\dot\var\sim 0.5\cdot
10^{14}\fr{\rm{erg}}{\rm{cm}^3 \rm{s}},$$ gives the following lower
bounds for the D-dimensional Plank mass: $M_D\sim 1.3 \rm{TeV}$ for
$d=2, M_D\sim 1.09\cdot 10^{-2} \rm{TeV}$ for $d=3$ and $M_D\sim
3.8\cdot 10^{-4} \rm{TeV}$ for d=4. Thus, the resonant conversion
plasmon-graviton in magnetic stars has to be included into the list
of basic mechanisms leading to astrophysical restrictions on the
parameters od the ADD model \cite{BHKZ99}-\cite{HR03}.
\section{Conclusion}
We presented the calculation of the energy-loss rate  due to the
emission of KK gravitons via plasmon-graviton transformation
processes in the magnetized non-relativistic plasma. The calculation
was made by using the kinetic approach taking into account the
collective effects.  The estimates show that this process can
compete with other mechanisms of the KK dissipation in magnetic
stars such as bremsstrahlung.

 The work was supported by RFBR project 11-02-01371-a.

\end{document}